\documentclass[aps,prl,twocolumn,groupedaddress,nofootinbib,preprintnumbers]{revtex4}

\usepackage{graphicx}
\usepackage{dcolumn}
\usepackage{bm}
\usepackage{latexsym}
\usepackage{amsfonts}
\usepackage{amssymb}
\usepackage{amsmath}
\usepackage{verbatim}
\usepackage{xcolor}

\begin{document}

\preprint{IPMU12-0123}

\title{Anisotropic Friedmann-Robertson-Walker universe from nonlinear massive gravity}

\author{A. Emir G\"umr\"uk\c{c}\"uo\u{g}lu}
\affiliation{Kavli Institute for the Physics and Mathematics of the Universe, Todai Institutes for Advanced Study, University of Tokyo, 5-1-5 Kashiwanoha, Kashiwa, Chiba 277-8583, Japan}

\author{Chunshan Lin}
\affiliation{Kavli Institute for the Physics and Mathematics of the Universe, Todai Institutes for Advanced Study, University of Tokyo, 5-1-5 Kashiwanoha, Kashiwa, Chiba 277-8583, Japan}

\author{Shinji Mukohyama}
\affiliation{Kavli Institute for the Physics and Mathematics of the Universe, Todai Institutes for Advanced Study, University of Tokyo, 5-1-5 Kashiwanoha, Kashiwa, Chiba 277-8583, Japan}

\date{\today}

\begin{abstract}
 Motivated by the recent no-go result of homogeneous and isotropic
 solutions in the nonlinear massive gravity, we study fixed points of
 evolution equations for a Bianchi type--I universe. We find a new
 attractor solution with non-vanishing anisotropy, on which the physical
 metric is isotropic but the St\"uckelberg configuration is
 anisotropic. As a result, at the background level, the solution
 describes a homogeneous and isotropic universe, while a statistical
 anisotropy is expected from perturbations, suppressed by smallness of
 the graviton mass.
\end{abstract}

\maketitle

{\bf Introduction.~}
General relativity, which describes long range gravitational
interactions, is in agreement with current experimental and observational
data. On the other hand, from a theorist's point of view, it is an
interesting question whether the range of gravity can be consistently
made to be finite, or equivalently, whether graviton can have a non-zero
mass. This question has also phenomenological relevance, since the
graviton mass may introduce new terms that mimic dark energy and thus
may source the late-time acceleration of the universe.

Assigning a mass to graviton has been one of the most challenging problems in classical field theory for the last 70 years. The linear theory of Fierz and Pauli \cite{Fierz:1939ix} gives rise to a discontinuity in the observables \cite{vdvz}, which can be alleviated by nonlinear terms \cite{Vainshtein:1972sx}. However, such terms generically introduce the so-called Boulware-Deser (BD) ghost \cite{Boulware:1973my}, spoiling the stability of the theory.

Only recently, a nonlinear extension of the massive gravity theory has
been introduced \cite{deRham:2010ik, deRham:2010kj}, where the BD ghost
is systematically removed by construction. The theoretical and
phenomenological possibilities brought this theory a significant
attention.

The nonlinear massive gravity allows self-accelerating open
Friedmann-Robertson-Walker (FRW) solutions with the Minkowski fiducial
metric \cite{Gumrukcuoglu:2011ew} as well as flat/closed/open FRW
solutions with general FRW fiducial metric
\cite{Gumrukcuoglu:2011zh}. Unlike the other branch of solutions
\cite{Fasiello:2012rw,Langlois:2012hk}, these backgrounds evade the
Higuchi bound \cite{Higuchi:1986py} and thus are free from ghost at the
linearized level even when the expansion rate is significantly higher
than the graviton mass. This is because there are only two propagating
modes on these backgrounds. However, these constructions exhibit a ghost
instability at nonlinear order in perturbations
\cite{DeFelice:2012mx}. This is a consequence of the FRW symmetries; in
order to obtain a stable solution, some of these symmetries need to be
broken.

An inhomogeneous background solution was obtained in
\cite{D'Amico:2011jj}, where the observable universe is approximately
FRW for a horizon size smaller than the Compton length of
graviton. Similar solutions with inhomogeneities in the St\"uckelberg
sector, meaning that the physical metric and the fiducial metric do not
have common isometries acting transitively, were found in
\cite{brokenFRW}. Note that those inhomogeneous solutions cannot be
isotropic everywhere since isotropy at every point implies
homogeneity~\cite{Wald_GR}. Note also that cosmological perturbations
can in principle probe inhomogeneities in the St\"uckelberg sector. For
example, generic spherically-symmetric solutions are isotropic only when
they are observed from the center of the universe.

The goal of the present paper is to introduce an alternative option,
where the assumption of isotropy is dropped but homogeneity,
i.e. the cosmological principle, is kept. In a region with relatively
large anisotropy, we find an attractor solution. On the attractor, the
physical metric is still isotropic, and the background geometry is of
FRW type. Hence, the thermal history of the standard cosmology can be
accommodated in this class of solutions. However, the St\"uckelberg
field configuration is anisotropic, which may lead to effects at the
level of the perturbations, suppressed by smallness of the graviton
mass.

\begin{figure}
\includegraphics[scale=0.4]{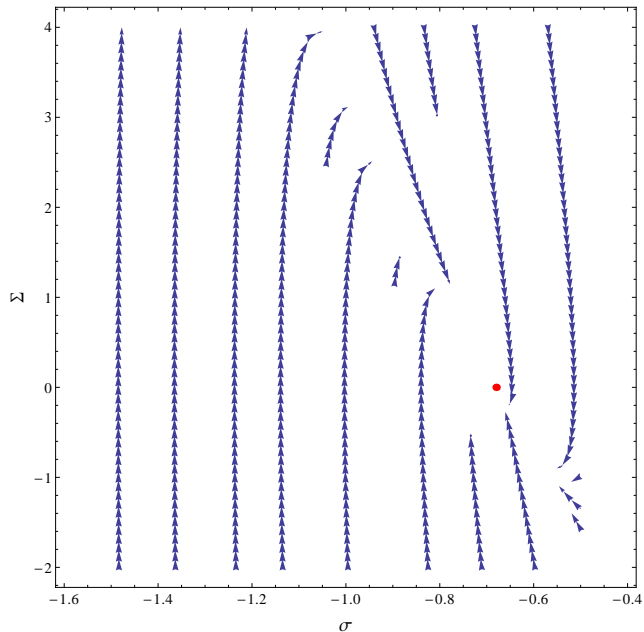}
\caption{\label{fig:flow} The phase flow for ($\sigma$, $\Sigma$) for parameters ($\alpha_3$, $\alpha_4$, $\lambda$, $\mu$) =($-0.05,\,1,\,0,\,20$). The flow is directed toward the red dot at ($\sigma$, $\Sigma$) = ($0.5$, $0$), which is the fixed point obtained by solving Eqs.(\ref{eq1})-(\ref{eq3b}).}
\end{figure}

{\bf The action and background.~}
We consider a simple description of the universe at present time. We
assume that the late-time acceleration is sourced by a cosmological
constant $\Lambda$, as well as the contribution from the graviton
mass. (Setting $\Lambda=0$ corresponds to self-accelerating solutions as
in the example shown in Fig.\ref{fig:flow}.) For this purpose, the
vacuum configuration is sufficient. The action with the graviton mass
term, constructed by imposing the absence of BD ghost in the decoupling
limit \cite{deRham:2010kj}, is
\begin{equation}
I = \tfrac{M_{pl}^2}{2}\int d^4x\sqrt{-g}[R-2\,\Lambda+m_g^2({\cal L}_2+\alpha_3{\cal L}_3+\alpha_4{\cal L}_4)]\,,
\end{equation}
where
\begin{align}
 {\cal L}_2 & =  [{\cal K}]^2-[{\cal K}^2]\,, \nonumber\\
 {\cal L}_3 & =  \tfrac{1}{3}
  ([{\cal K}]^3-3[{\cal K}][{\cal K}^2]+2[{\cal K}^3]),
  \nonumber\\
 {\cal L}_4 & =\tfrac{1}{12}
  ([{\cal K}]^4-6[{\cal K}]^2[{\cal K}^2]+3[{\cal K}^2]^2
   +8[{\cal K}][{\cal K}^3]-6[{\cal K}^4]).
\end{align}
In the above, the square brackets denote trace operation and
\begin{equation}
{\cal K}^\mu _\nu = \delta^\mu _\nu
 - \left(\sqrt{g^{-1}f}\right)^{\mu}_{\ \nu}\,.
\label{Kdef}
\end{equation}
Here, $g_{\mu\nu}$ is the physical metric, while the space-time tensor
$f_{\mu\nu}$ is the fiducial metric, whose vacuum expectation value
gives rise to the breaking of general coordinate invariance.

For the physical metric, we adopt the axisymmetric Bianchi type--I metric, which is the simplest anisotropic extension of FRW ansatz
\begin{equation}
 g^{(0)}_{\mu\nu}dx^{\mu}dx^{\nu} =
  -N^2dt^2 + a^2
  [e^{4\sigma}dx^2+e^{-2\sigma}\delta_{pq}dy^idy^j],
\end{equation}
where $N$, $a$ and $\sigma$ are functions of $t$, Greek indices denote
space-time coordinates, while $i,j=2,3$. As for the fiducial metric, we
assume the flat FRW form as
\begin{equation}
 f_{\mu\nu} =
  -n^2\partial_{\mu}\phi^0\partial_{\nu}\phi^0
  +\alpha^2(\partial_\mu\phi^1\,\partial_\nu \phi^1+\delta_{ij}\partial_{\mu}\phi^i\partial_{\nu}\phi^j),
\label{fiducial}
\end{equation}
where $n$ and $\alpha$ are functions of the temporal St\"uckelberg field $\phi^0$.
The form (\ref{fiducial}) includes a de Sitter fiducial as a special case, with $H_f \equiv \dot{\alpha}/\alpha n = {\rm constant}$.

Varying the St\"uckelberg fields around the background value $\phi^a = x^a + \pi^a$, the variation of the mass term up to first order is
\begin{equation}
 I = I^{(0)} + M_{Pl}^2m_g^2
  \int d^4xNa^3n\pi^0 {\cal E}_{\phi} + O[(\pi^a)^2],
\label{eqstuck}
\end{equation}
giving the equation of motion
\begin{align}
{\cal E}_\phi \equiv &J_\phi^{(x)}\,\left(H+2\,\Sigma -H_f\,e^{-2\,\sigma}\,X\right)
\nonumber\\
&\qquad\qquad+2\,J_\phi^{(y)}\,\left(H-\Sigma-H_f\,e^{\sigma}\,X\right)=0\,,
\label{eq:stuckfieldeq}
\end{align}
where

\vspace{-.5cm}%
\begin{align}
J_\phi^{(x)}  &\equiv \gamma_1-2\,\gamma_2\,e^\sigma\,X+\gamma_3\,e^{2\sigma}\,X^2\,,\nonumber\\
J_\phi^{(y)}  &\equiv \gamma_1-\,\gamma_2\,(e^{-2\sigma}+e^\sigma)\,X+\gamma_3\,e^{-\sigma}\,X^2\,,
\end{align}
with
$\gamma_1\equiv3+3\,\alpha_3+\alpha_4$, $\gamma_2\equiv1+2\,\alpha_3+\alpha_4$, $\gamma_3\equiv\alpha_3+\alpha_4$, $H \equiv \tfrac{\dot{a}}{a\,N}$, $\Sigma \equiv \tfrac{\dot{\sigma}}{N}$ and $X\equiv \tfrac{\alpha}{a}$. The expansion rate for the fiducial metric $H_f$ is related to the invariants of the field space metric, and is independent of the choice of the background values of $\phi^a$. Thus, Eq.(\ref{eq:stuckfieldeq}) can be interpreted as an algebraic equation for $\alpha$ (or equivalently for $X$), instead of a differential equation.

Varying the action with respect to $g_{\mu\nu}$, the field equations for the physical metric are obtained as
\begin{widetext}
\begin{align}
3\left(H^2 - \Sigma^2\right)-\Lambda &= m_g^2\left[
-(3\,\gamma_1-3\,\gamma_2+\gamma_3)+
 \gamma_1\,(2\,e^\sigma +e^{-2\sigma})X- \gamma_2(e^{2\sigma}+2\,e^{-\sigma})\,X^2 +\gamma_3\,X^3\right],\nonumber\\
\frac{\dot{\Sigma}}{N}+3H\Sigma &=
\tfrac{m_g^2}{3} (e^{-2\,\sigma}-e^\sigma)X\left[ \gamma_1 -\gamma_2(e^\sigma+r)X+ \gamma_3\,r e^\sigma X^2\right]\,,
\label{eqeins}
\end{align}
\end{widetext}
where
\begin{equation}
r \equiv \frac{n\,a}{N\,\alpha} = \frac{1}{X\,H_f}\,\left(\frac{\dot{X}}{N\,X}+H\right)\,.
\end{equation}
Additionally, there is also an equation for $\dot{H}$, which can be recovered by combining Eq.(\ref{eq:stuckfieldeq}) with Eq.(\ref{eqeins}).

{\bf Fixed Points.~}
We consider a de Sitter fiducial metric ($H_f=const.$) and seek
solutions with $\dot{H}=\Sigma=\dot{X}=0$.
The constancy of $X$ allows us to express $H$ as $H= H_f\,X\,r$.
In this setup, the independent equations become
\begin{widetext}
\begin{align}
&3\lambda-(3\gamma_1-3\gamma_2+\gamma_3)+\gamma_1(2e^{\sigma}+e^{-2\sigma})X-\left[\gamma_2(2e^{-\sigma}+e^{2\sigma})+3\,r^2\mu^{-2}\right]X^2+\gamma_3X^3 = 0,
\label{eq1}\\
&(e^{\sigma}-1)\left[\gamma_1-\gamma_2(r+e^\sigma)X+\gamma_3 e^{\sigma}rX^2\right] =  0,
\label{eq2}\\
&\gamma_1(3r-2e^\sigma-e^{-2\sigma})-2\gamma_2\left[(2e^{\sigma}+e^{-2\sigma})r -(e^{2\sigma}+2e^{-\sigma})
 \right]X+\gamma_3\left[(e^{2\sigma}+2e^{-\sigma})r-3\right]X^2   =  0,
\label{eq3b}
\end{align}
\end{widetext}
where $\lambda\equiv \tfrac{\Lambda}{3m_g^2}$ and $\mu \equiv \tfrac{m_g}{H_f}$  are dimensionless parameters.

For $\sigma=0$, the set of equations is reduced to that for isotropic configurations, which was already investigated in \cite{Gumrukcuoglu:2011ew, Gumrukcuoglu:2011zh}. Assuming $\sigma\ne 0$ and using (\ref{eq2}), Eq.(\ref{eq3b}) can be rewritten as
\begin{equation}
 (\gamma_1-\gamma_2 Xe^{\sigma})(e^{\sigma}-r)
  (re^{2\sigma}-1) = 0\,.
  \label{eq3}
\end{equation}
Considering (\ref{eq3}) as an algebraic equation for $e^\sigma$, there are three solutions:
\begin{equation}
e^\sigma = \left\{\frac{\gamma_1}{\gamma_2\,X}\,,\; r\,,\;r^{-1/2}\right\}\,.
\label{sigsols}
\end{equation}
We now consider each solution separately.

\underline{\it Case I.} $e^\sigma =
\tfrac{\gamma_1}{\gamma_2\,X}$.~~ Using this solution in
Eq.(\ref{eq2}) gives $X = \gamma_1/\gamma_2$, implying $\sigma=0$.
In other words, this solution is isotropic and thus is not of our
interest.

\underline{\it Case II.} $e^\sigma = r$.~~ In this case,
Eq.(\ref{eq2}) gives
\begin{equation}
(r-1)\left[\gamma_1 -2\,\gamma_2\,r\,X + \gamma_3\,(r\,X)^2\right]=0\,.
\end{equation}
This equations have two solutions; The first solution is $r=1$,
and leads to isotropy $\sigma =0$ which is not our interest.
The second solution gives $rX=(\gamma_2\pm
\sqrt{\gamma_2^2-\gamma_1\gamma_3})/\gamma_3$, which reduces
Eq.(\ref{eq1}) to a nontrivial constraint between the parameters
of the theory. Since this case requires a fine-tuning of a
parameter, it is not of our interest either.

\underline{\it Case III.} $r=e^{-2\,\sigma}$.~~ With this
solution, Eq.(\ref{eq2}) is reduced to
\begin{equation}
\gamma_1 e^{\sigma} - \gamma_2 (e^{2 \sigma}+e^{- \sigma})X + \gamma_3 X^2 = 0.
\label{intermediate}
\end{equation}
while Eq.(\ref{eq1}) becomes
\begin{align}
(3\lambda-3\gamma_1+3\gamma_2-\gamma_3)  + \gamma_1(e^{-2\sigma}+2e^{\sigma})X  \quad&\nonumber\\
-[ \gamma_2(2e^{-\sigma}+e^{2\sigma})
    + 3\,e^{-4\sigma}\,\mu^{-2}]X^2 +\gamma_3X^3 &=  0.
\end{align}
Combining these two equations, we obtain an expression linear in $X$,
\begin{equation}
 X = \frac{3\gamma_1+[\gamma_1 \gamma_2-\gamma_3^2+3 \gamma_3(\gamma_2-\gamma_1+\lambda)]\mu^2e^{3\sigma}}
  {(e^{\sigma}+e^{-2\sigma})\,[3\gamma_2+(\gamma_2^2-\gamma_1\gamma_3)]\mu^2e^{3\sigma}}.
  \label{eqn:X-for-caseIII-2}
\end{equation}
and an equation which only depends on $\sigma$
\begin{equation}
c_0+c_1 e^{3\sigma}+ c_2e^{6\sigma} + c_3e^{9\sigma} = 0,
 \label{eqn:r-for-caseIII-2}
\end{equation}
where
\begin{widetext}
\begin{eqnarray}
 c_0 & = &
3\gamma_2\left(\gamma_1^2+3\gamma_2^2-3\gamma_1\gamma_2-\gamma_2\gamma_3+3\gamma_2\lambda\right)\mu^2-9\gamma_1^2 \,,  \nonumber\\
 c_1 & = & (\gamma_2^2 -\gamma_1\gamma_3)\,\left[-6(3\gamma_1-3\gamma_2+\gamma_3-3\lambda)+\left(\gamma_1^2+3\gamma_2^2-3\gamma_1\gamma_2-\gamma_2\gamma_3+3\gamma_2\lambda\right)\mu^2\right]\mu^2\,,
  \nonumber\\
 c_2 & =& \left[3\gamma_2+(2\gamma_2^2-2\gamma_1\gamma_3)\mu^2\right]\left(\gamma_1^2+3\gamma_2^2-3\gamma_1\gamma_2-\gamma_2\gamma_3+3\gamma_2\lambda\right)\mu^2-(\gamma_1\gamma_2-3\gamma_1\gamma_3+3\gamma_2\gamma_3-\gamma_3^2+3\gamma_3\lambda)^2\mu^4\,,
 \nonumber\\
 c_3 & = &
  (\gamma_2^2-\gamma_1\gamma_3)\left(\gamma_1^2+3\gamma_2^2-3\gamma_1\gamma_2-\gamma_2\gamma_3+3\gamma_2\lambda\right)\mu^4\,,
\end{eqnarray}
\end{widetext}
For a given set of parameters $(\alpha_3,\alpha_4,\lambda,\mu)$, one can solve the cubic equation (\ref{eqn:r-for-caseIII-2}) for
$e^{3\sigma}$ and then use (\ref{eqn:X-for-caseIII-2}) to
calculate the corresponding value of $X$.

{\bf Local Stability.~}
We now introduce homogeneous perturbations around the fixed point described in the third case above.
\begin{align}
 H&= H_f [r_0X_0 + \epsilon \,h_1(t) + O(\epsilon^2)], \nonumber\\
  \sigma &= \sigma_0 + \epsilon \,\sigma_1(t) + O(\epsilon^2), \nonumber\\
  X &= X_0 + \epsilon \,X_1(t) + O(\epsilon^2),
\end{align}
where ($X_0$, $\sigma_0$, $r_0=e^{-2\sigma_0}$) is the background
representing the fixed point. Adopting this expansion, we calculate the equations of motion up to ${\cal O}(\epsilon)$. At linear order, $\sigma_1$ can be decoupled from the remaining ${\cal O}(\epsilon)$ quantities, and a second-order evolution equation is obtained as
\begin{equation}
 \frac{d^2\sigma_1}{d\tau^2}
  + 3X_0 e^{-2\sigma}\frac{d\sigma_1}{d\tau}
  + M^2\sigma_1 = 0,
\end{equation}
where
\begin{align}
 M^2 & =  \frac{X_0^2 \mu^2 e^{-4\sigma_0}}{2}\,\left(\frac{d_1\,(3\,d_1-d_2)(6+d_1\,\mu^2)}{2\,d_2-d_1{^2}\,\mu^2}\right)\,,\nonumber\\
d_1 &\equiv (e^{3\,\sigma_0}-1)\,\left[\gamma_2 -\gamma_3 e^{\sigma_0}\,X_0\right]\,,
\nonumber\\
d_2 &\equiv (e^{3\,\sigma_0}-1)\,\left[\gamma_2(3 +2\,e^{3\,\sigma_0})-5\,\gamma_3e^{\sigma_0}\,X_0\right]\,,
\end{align}
while the dimensionless time coordinate $\tau$ is defined by $ d\tau = H_fNdt$.
The fixed point is locally stable if
\begin{equation}
 M^2 > 0.
\label{local}
\end{equation}

{\bf Global Stability.~}
To study the global stability of the fixed point, we consider an example with
\begin{equation}
\lambda=0 \,,\quad \mu = 20 \,,\quad \alpha_3= -1/20\,,\quad \alpha_4=1\,,
\label{params}
\end{equation}
for which the local stability condition (\ref{local}) is
satisfied. For this parameter set, there is only one set of real
solutions to the equations of motion (\ref{eq1})-(\ref{eq3b})
\begin{equation}
X\simeq 4\,, \quad
e^{\sigma} \simeq \tfrac{1}{2}\,,\quad
r \simeq 4\,.
\label{example}
\end{equation}
In order to determine the phase flow, we first reduce the system of equations. Using Eq.(\ref{eq:stuckfieldeq}), the first of Eq.(\ref{eqeins}) and their time derivatives, we can express $X$, $H$ and their derivatives in terms of $\sigma$ and $\Sigma$. Since these equations are nonlinear, there are always more than one solution. For the parameter set (\ref{params}), we find that there are three branches of solutions which give $X>0$ and $H>0$. For each branch, we use this solution with the second of Eq.(\ref{eqeins}). As a result we obtain, for a set of $(\sigma, \Sigma)$, the corresponding set of $(\dot{\sigma},\dot{\Sigma})$ pair. Out of the three branches, only one contains an attractor. The flow corresponding to this branch is shown in Fig.\ref{fig:flow}. The flow proceeds towards the fixed point we found in Eq.(\ref{example}).

{\bf Discussion.}
The recently introduced nonlinear massive gravity theory
\cite{deRham:2010ik,deRham:2010kj} provides a new framework to address
some of the intriguing issues in theoretical physics and cosmology, such as
the possibility of finite range gravity and the mystery of dark
energy. Although the theory admits homogeneous and isotropic solutions,
these suffer from an unavoidable nonlinear ghost
\cite{DeFelice:2012mx} or a linear ghost \cite{Fasiello:2012rw}. Since
this is a consequence of the FRW symmetries, either homogeneity or
isotropy needs to be broken in order to obtain a stable solution.

In the present paper, for the first time in the nonlinear massive
gravity theory, we explored regions with relatively large anisotropy for
homogeneous attractor solutions. The classification of fixed points
revealed the existence of a single anisotropic attractor. The local and
global stability analyses indicate that, a universe with a sufficiently
large anisotropy at the onset of the late-time accelerated expansion
should flow to this point.

A very interesting implication is that the scale factors corresponding
to the two directions differ only by a constant normalization, thus the
expansion rate is completely isotropic. In general relativity, such a
solution will be identical to an isotropic universe, up to a coordinate
redefinition. Conversely, in nonlinear massive gravity, such a
redefinition cannot remove the anisotropy completely; it is merely
shifted from the physical metric to the fiducial metric. Although the
background metric is of FRW type, the signature from the
anisotropy will be imprinted on the spectrum of cosmological
perturbations. The statistical anisotropy signal is expected to be
suppressed by smallness of the graviton mass $m_g$. The type of
anisotropy, i.e. statistical anisotropy for perturbations without
anisotropic background expansion, is totally new. For example, none of
the anisotropic inflation scenarios \cite{Soda:2012zm} has this type of
anisotropy. Detailed analysis of perturbations and comparison with
observational data are worthwhile pursuing. As the first step, a
preliminary analysis of perturbations indicates that the anisotropic
attractor solutions found in this paper are free from ghost for a range
of parameters \cite{glp4}.

As a sensible effective theory, the theory can be reliable only below a cutoff scale $\Lambda_n =(M_{Pl}\,m_g^{n-1})^{1/n}$, which is much larger than the graviton mass $m_g$ for $n>0$. The exact form of this scale will be determined by a detailed analysis of perturbations in a future publication \cite{glp4}. For the scenario discussed in the present work, we associated the late time acceleration with the graviton mass term, thus the present expansion rate is taken to be of the order of the mass term. For instance, for the case given in Eqs.~(\ref{params}-\ref{example}), the constant expansion rate on the fixed point is $H = 0.8 \,m_g$. This is clearly well below the order of the cutoff scale $\Lambda_n$ for $n>0$.

~~~
\begin{acknowledgments}
This work was supported by the World Premier International Research Center Initiative (WPI Initiative), MEXT, Japan. S.M.\ also acknowledges the support by Grant-in-Aid for Scientific Research 24540256 and 21111006, and by Japan-Russia Research Cooperative Program.
\end{acknowledgments}

\end{document}